\documentclass[15pt a4paper]{article}

\usepackage{latexsym}
\usepackage{amsmath}
\usepackage{amssymb}
\usepackage{graphicx}
\graphicspath{{./}{./figs/}}

\begin{document}

\title{\bf On-off intermittency in small-world networks of chaotic maps \thanks{IEEE International Symposium on Circuits and Systems, 2005 (ISCAS 2005).
Vol. 1, pp. 288-291, 23-26 May 2005, Kobe, Japan.}}
\author{Chunguang Li$^1$\thanks{Corresponding author, Email: cgli@uestc.edu.cn}, and Jin-Qing Fang$^2$}
\date{\small $^1$Centre for Nonlinear and Complex Systems, School of Electronic
Engineering, \\University of Electronic Science and
Technology of China, Chengdu, 610054, P. R. China.\\
$^2$China Institute of Atomic Energy, P. O. Box 275-81, Beijing,
102413, P. R. China } \maketitle
\begin{abstract}
Small-world networks are highly clustered networks with small
average distance among the nodes. There are many natural and
technological networks that present this kind of connections. The
on-off intermittency is investigated in small-world networks of
chaotic maps in this paper. We show how the small-world topology
would affect the on-off intermittency behavior. The distributions
of the laminar phase are calculated numerically. The results show
that the laminar phases obey power-law distributions.
\end{abstract}
A great deal of research interest on the theory and applications
of small-world networks have been aroused [1] since the pioneering
work of Watts and Strogatz [2]. Some common properties of complex
networks, such as the Internet, power grids, forest fires, and
disordered porous media, are mainly determined by the way of
connections among their vertices or occupied sites. One extremal
case is a regular network that has a high degree of local
clustering and a large average distance, while the other extremal
case is a random network with negligible local clustering and a
small average distance. In between, a small-world network is a
special case of complex networks with a high degree of local
clustering as well as a small average distance. Recently, the
dynamics of small-world networks were studied in different fields.
For example, many authors have investigated the synchronization of
small-world networks of phase oscillators, coupled map lattices
and general dynamical systems [3]. Bifurcation, fractals and chaos
in small-world networks were studied in [4]. Self-organized
criticality on small-world networks was studied in [5]. Turbulence
in small-world networks was investigated in [6]. Oscillators death
phenomenon on small-world networks was reported in [7]. Stochastic
resonance in small-world networks was studied in [8]. And
dynamical behaviors of small-world neural networks were
investigated in [9].

Recently, the so called on-off intermittency has been reported in
literatures [10-12]. This type of intermittency is characterized
by a two state nature. The ``off" state, which is nearly constant,
and remains so for long periods of time and is suddenly changed by
a burst, the so called ``on" state, which departs quickly, and
return quickly to the ``off" state.  Moreover, the power-law
distributions of the laminar phases was also observed and
discussed in these literatures. In [10], on-off intermittency in
low-dimensional systems were studied. In [11, 12], the authors
studied on-off intermittencies in {\it regular} coupled systems.
In [11], on-off intermittency was investigated in a {\it
nearest-neighbor} coupled-map lattice by applying noise at a
single node. In [12], on-off intermittency in {\it globally}
coupled map lattices were studied.

However, connection topology in real-world networks are usually
not completely regular. And it is well-known that usually topology
structure affects network dynamics critically. It is interesting
to investigate how small-world topology would affect the on-off
intermittency of coupled chaotic map networks. In this paper, we
study this topic numerically. We fix the coupling coefficient to a
constant so that the globally coupled lattice is synchronous. Then
we decrease the connection-adding probability gradually. We found
when the connection-adding probability slightly less than a
critical value, the synchronous chaos is no longer stable and
on-off intermittency appears. When further decrease this
connection-adding probability, the intermittent dynamics is
eventually replaced by fully developed asynchronous chaos.

The network model of $N$ coupled chaotic maps studied in this
paper is described by the following equations:
\begin{equation}
x_i(t+1)=(1-\epsilon)f(x_i(t))+\frac{\epsilon}{N_i}\sum_{j=1}^Na_{ij}f(x_j(t))
\end{equation}
where $f(x)$ is a chaotic map, $\epsilon>0$ is the coupling
coefficient, $N_i$ is the number of maps connected to the $i$th
map. The matrix $A=\{a_{ij}\}$ encodes the connection topology: if
there is a connection between maps $i$ and $j$, $a_{ij}=a_{ji}=1$;
otherwise, $a_{ij}=0$. $a_{ii}=0$ for all $i$, and
$N_i=\sum_{j=1}^Na_{ij}$. We are interested in small-world
connection topology of network (1) in this paper.

The original small-world (SW) model introduced in [2] can be
described as follows. Take a one-dimensional lattice of $N$
vertices arranged in a ring with connections only between nearest
neighbors. We ``rewire'' each connection with probability $p$.
Rewiring in this context means reconnecting randomly the whole
lattice, with the constraint that no two different vertices can
have more than one connection in between, and no vertex can have a
connection with itself.

Note, however, that there is a possibility for the SW model to be
broken into unconnected clusters. This problem can be resolved by
a slight modification of the SW model, suggested by Newman and
Watts (NW) [13]. In the NW model, one does not break any
connection between any two nearest neighbors. Instead, one adds
with probability $p$ a connection between each unconnected pair of
vertices. Likewise, one does not allow a vertex to be coupled to
another vertex more than once, or to couple with itself. For
$p=0$, it reduces to the originally nearest-neighbor coupled
network; for $p=1$, it becomes a globally coupled network. From a
coupling-matrix point of view, network (1) with small-world
connections will evolve according to the rule that, in the
nearest-neighbor coupling matrix $A$, if $a_{ij}=0$, then set
$a_{ij}=a_{ji}=1$ with probability $p$.

In [12], the authors showed, for globally coupled networks, there
is a critical value $\epsilon_c$ of the coupling coefficient
$\epsilon$. When $\epsilon>\epsilon_c$, the network is
synchronous, and when $\epsilon$ slightly less than $\epsilon_c$,
on-off intermittency appears. In this paper, we are interested in
the NW small-world connection topology of network (1). It is
natural to ask {\it if we let $\epsilon>\epsilon_c$, whether
on-off intermittency can occur by decreasing the probability $p$
so as to make the network be a small-world.} In this paper, we
study this topic and provide a positive answer to this question.

In order to discuss the behavior of on-off intermittency, we will
work in the variable difference space
\begin{equation}
\Delta x_i(t)=x_i(t)-\bar{x}(t)=x_i(t)-\frac{1}{N}\sum_{j=1}^N
x_j(t),\hspace{0.5cm} i=1,\cdots,N
\end{equation}
By subtracting the average, all variable differences $\Delta x_i$
keep their desynchronized parts and the synchronous chaos is
eliminated.

\begin{figure}[htb]
\centering
\includegraphics[width=10cm]{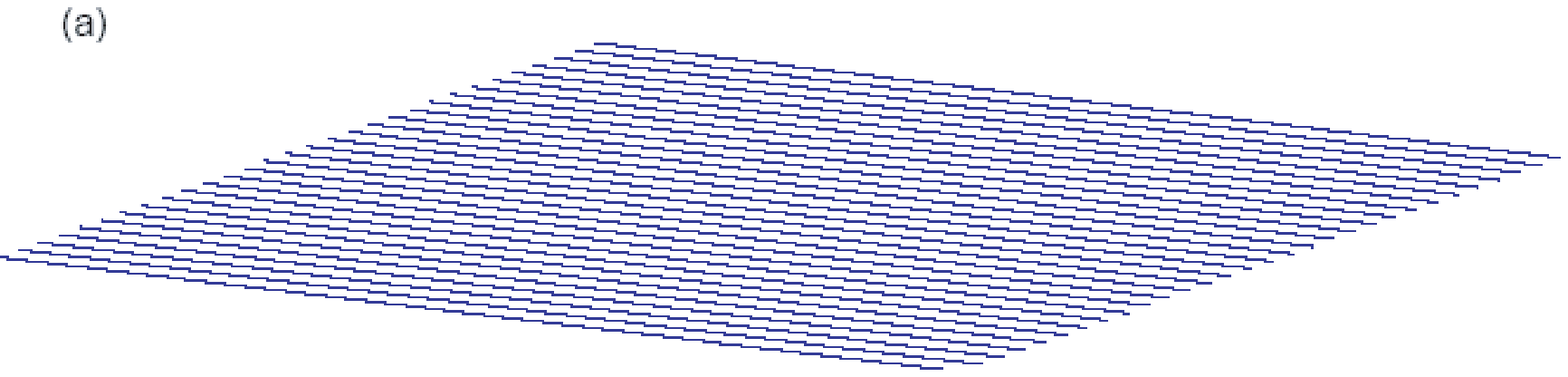}
\includegraphics[width=10cm]{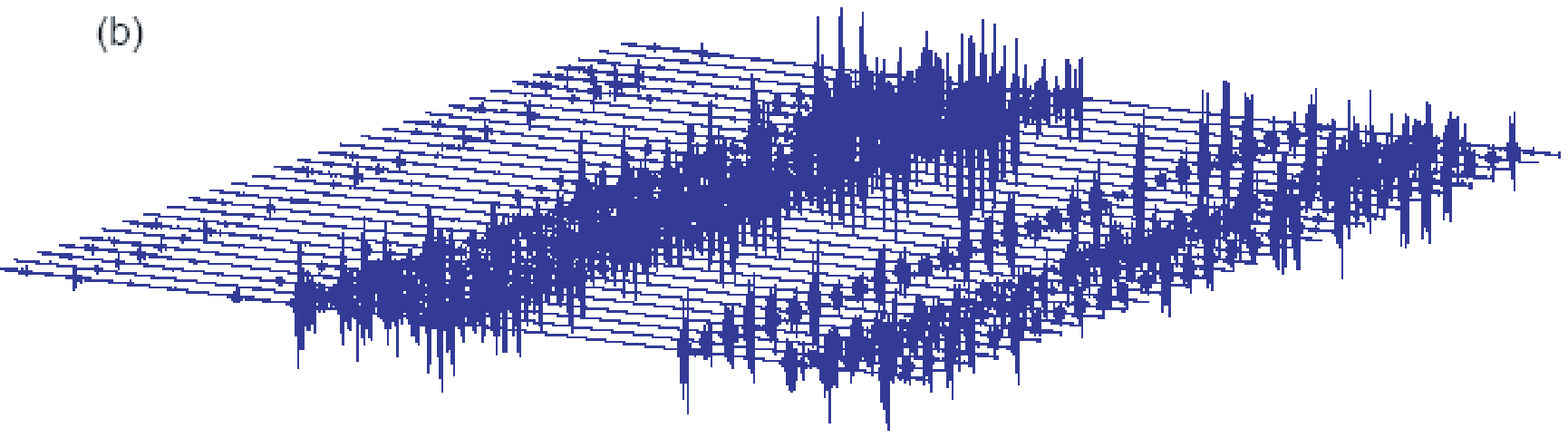}
\includegraphics[width=10cm]{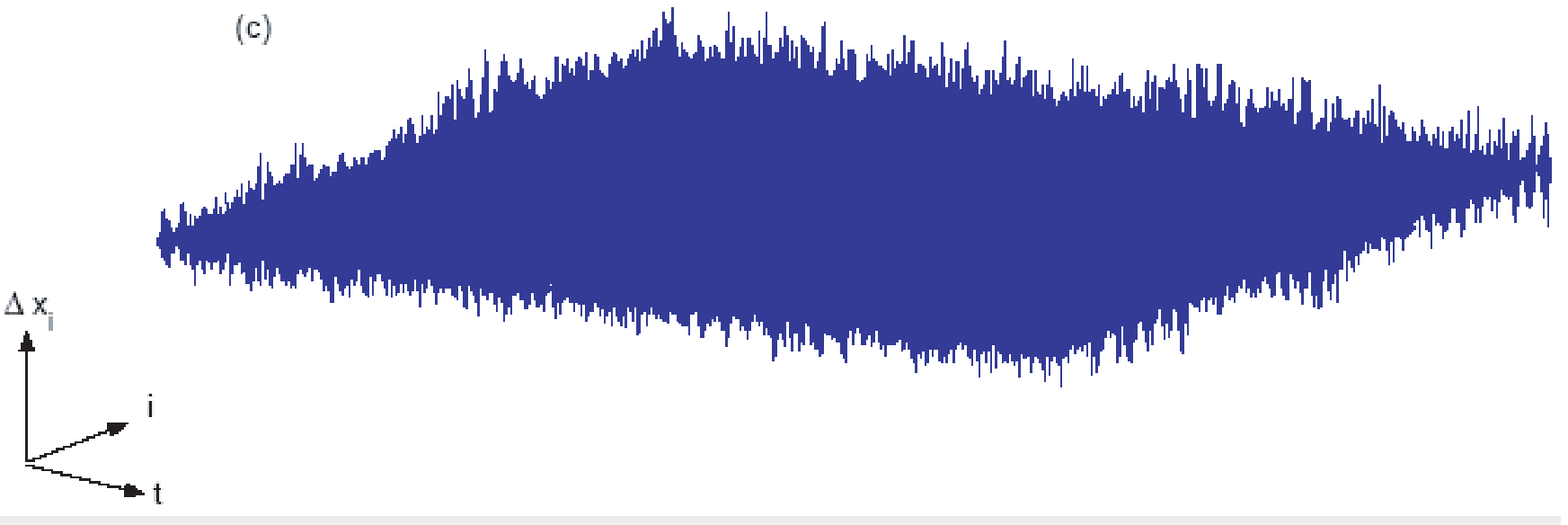}
\caption{The time series of $\Delta x_i(t)$ of 30 randomly
selected maps. (a) $p=0.3$, synchronous; (b) $p=0.27$,
intermittency; and (c) $p=0.12$, asynchronous chaos.}
\end{figure}

\begin{figure}[htb]
\centering
\includegraphics[width=10cm]{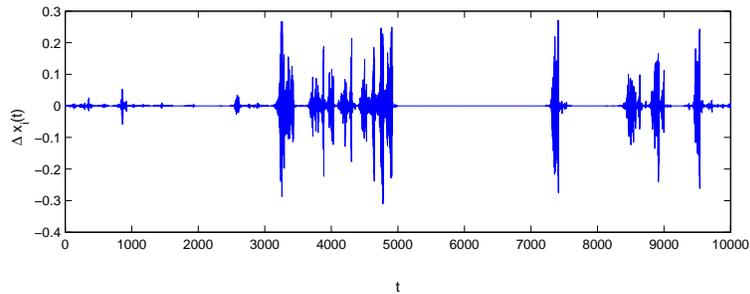}
\caption{The time series of $\Delta x_i(t)$ of a randomly selected
map $i$ for $p=0.27$}
\end{figure}
\begin{figure}[htb]
\centering
\includegraphics[width=7cm]{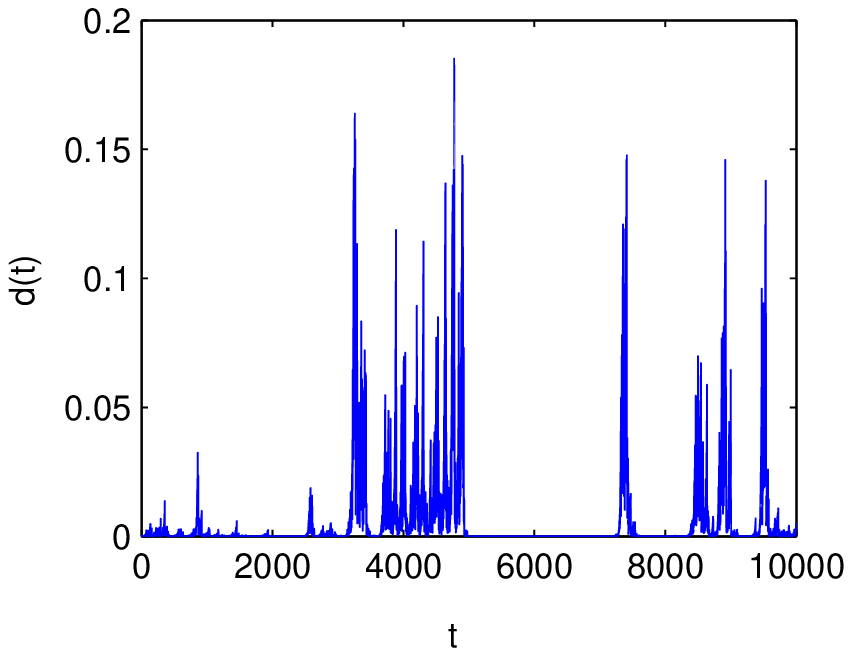}
\caption{The time series of $d(t)$ for $p=0.27$}
\end{figure}

For simplicity, we consider only 1-D chaotic maps in this paper.
Let $f(x)=1-ax^2$ be the logistic map. For $a=1.9$, this map is
chaotic. Consider a globally coupled network of $N=100$ such maps.
From [12], we can get a critical value $\epsilon_c=0.4225$. When
$\epsilon>\epsilon_c$, the globally coupled network is
synchronous. In the following we let $\epsilon=0.6$. Obviously,
with this coupling coefficient, the globally coupled network is
synchronous. We gradually decrease the connection-adding
probability $p$. We found when $p$ less than a critical value
$p_c\approx 0.29$, the synchronous states lose their stability.
Fig.1 (a) shows the time evolution dynamics of $\Delta x_i(t)$ of
$30$ randomly selected maps for $p=0.3>p_c$. For the 30 maps
displayed in the figure, we observe synchronized chaos in which
the plotted quantity is uniformly zero. For $p=0.27$, which is
slightly below $p_c$, the time evolution dynamics of $\Delta
x_i(t)$ of $30$ randomly selected maps is shown in Fig.1 (b),
which is interspersed with bursts away from the synchronization
attractor, suggesting the occurrence of on-off intermittency. This
intermittent dynamics is eventually replaced by fully developed
asynchronous chaos as the probability $p=0.12$ is far removed from
the critical value, as shown in Fig. 1 (c). In Fig. 2, we show the
intermittent dynamics of $\Delta x_i(t)$ of a randomly selected
map $i$ of the network for $p=0.27$. Further, we define a variable
to measure the distance between the system state and the
synchronization manifold as follows [12]:
\begin{equation}
d(t)=\frac{1}{N}\sum_{j=1}^N|x_j(t)-\bar{x}(t)|
\end{equation}
Fig. 3 show the value of $d$ as a function of time $t$ for
$p=0.27$, which is also a on-off intermittent time series.

A commonly used characteristic of on-off intermittent time series
is the laminar length distribution. Let $\tau$ denote the
threshold value of $|\Delta x_i|$ such that for $|\Delta
x_i|\geq\tau$ the signal is considered ``on" and for $|\Delta
x_i|<\tau$ the signal is considered ``off". The length of the
laminar phase is defined as the length of the off state. We use
$P_n$ to represent the probability of the laminar phase of length
$n$, namely, $P_n=M_n/N$, where $N$ is the total number of
segments of the laminar phase, and $M_n$ the number of those of
length $n$. We let $\tau=0.001$ and $p=0.27$ in our simulations.
We collect 10000 iterations for each map. We plot the numerically
calculated distribution for the length of laminar phase for 100
coupled logistic maps in Fig. 4, which is a power-law distribution
with a heavy-tail. In Fig. 5, we plot the distribution of laminar
length of the time series $d$. The time series used here is
constructed in the same way as that shown in Fig.3. In Fig. 6, we
plot the distribution of laminar length of the time series
$|\Delta x_i|$ of a randomly selected map $i$. As we can see from
these two figures, the distributions are also power-laws with
heavy-tails.

\begin{figure}[htb]
\centering
\includegraphics[width=7cm]{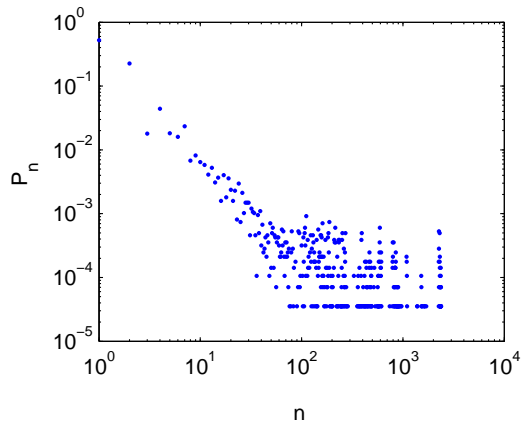}
\caption{The probability distribution $P_n$ of laminar length of
on-off intermittent time series for 100 maps.}
\end{figure}
\begin{figure}[htb]
\centering
\includegraphics[width=7cm]{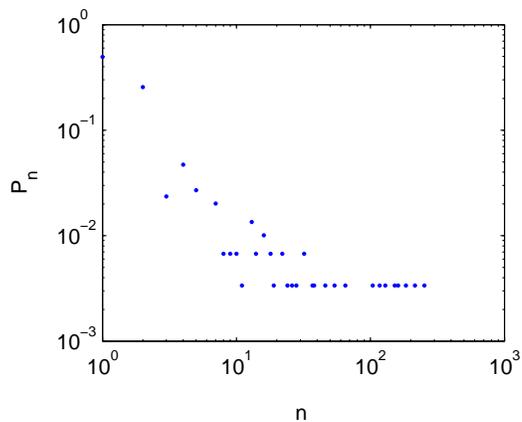}
\caption{The probability distribution $P_n$ of laminar length of
$d$ defined in Eq. (3).}
\end{figure}
\begin{figure}[htb]
\centering
\includegraphics[width=7cm]{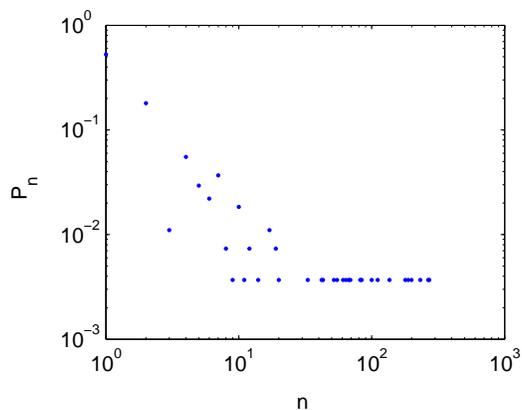}
\caption{The probability distribution $P_n$ of laminar length of
on-off intermittent time series for a randomly selected map.}
\end{figure}

In summary, in this paper, we have studied how the small-world
topology affects on-off intermittency of networks of coupled
chaotic maps. We found that with a fixed coupling coefficient
$\epsilon>\epsilon_c$, by decreasing the connection-adding
probability gradually, when $p$ slightly less than a critical
value, the synchronous chaos is no longer stable and on-off
intermittency appears. When further decrease $p$, the intermittent
dynamics is eventually replaced by fully developed asynchronous
chaos. The probability distributions of the length of laminar
phase are also presented to characterize the on-off intermittent
time series. As in may intermittent time series [10-12], the
length of laminar phases also obey power-law distributions. For
some other values of $N$ and $\epsilon>\epsilon_c$, and some other
chaotic maps, similar phenomena were also observed. We omitted
them.

This research was supported by the National Natural Science
Foundation of China under Grant 60271019, and the Youth Science
and Technology Foundation of UESTC under Grant YF020207.

\end{document}